\documentclass[letterpaper,twocolumn,10pt]{article}
\usepackage{usenix2019_v3}

\usepackage{stfloats}
\usepackage{amssymb,amsmath,amsthm}

\usepackage{array}
\usepackage{graphicx}
\usepackage{verbatim}
\usepackage{float}
\graphicspath{{figures-pdf/}}

\usepackage{multirow}
\usepackage{stfloats}
\usepackage{algorithm}
\usepackage{algorithmic}

\usepackage{calc}   

\newcommand{\tabincell}[2]{\begin{tabular}{@{}#1@{}}#2\end{tabular}}

\begin{document}

\date{}

\title{\Large \bf Adversarial Audio: A New Information Hiding Method and Backdoor for DNN-based Speech Recognition Models}

\author{
{\rm Yehao Kong, Jiliang Zhang}
\thanks{\textbf{Corresponding author.}}\\
College of Computer Science and Electronic Engineering\\
Hunan University, China\\
zhangjiliang@hnu.edu.cn
} 
\maketitle

\begin{abstract}
Audio is an important medium in people's daily life, hidden information can be embedded into audio for covert communication. Current audio information hiding techniques can be roughly classed into time domain-based and transform domain-based techniques. Time domain-based techniques have large hiding capacity but low imperceptibility. Transform domain-based techniques have better imperceptibility, but the hiding capacity is poor. This paper proposes a new audio information hiding technique which shows high hiding capacity and good imperceptibility. The proposed audio information hiding method takes the original audio signal as input and obtains the audio signal embedded with hidden information (called stego audio) through the training of our private automatic speech recognition (ASR) model. Without knowing the internal parameters and structure of the private model, the hidden information can be extracted by the private model but cannot be extracted by public models. We use four other ASR models to extract the hidden information on the stego audios to evaluate the security of the private model. The experimental results show that the proposed audio information hiding technique has a high hiding capacity of 48 cps with good imperceptibility and high security. In addition, our proposed adversarial audio can be used to activate an intrinsic backdoor of DNN-based ASR models, which brings a serious threat to intelligent speakers.
\end{abstract}

\section{Introduction}

With the rapid development of communication-related technologies, multimedia information such as image, audio, and video is generated in large quantities and brings great convenience to people. However, multimedia information services pose a potential threat to the legitimate rights of the information owner. As a different technology from traditional cryptography, information hiding techniques \cite{Anderson1996} that use the human's perceptual redundancy for digital signals and hide the secret information into the carrier to provide technical protection for the rights of multimedia information.

Since human auditory systems are more sensitive than visual systems, embedding secret information into audio media is more challenging than images. In addition, as an important medium in people's daily life communication, the audio has a good imperceptibility in the transmission of information and provides a lot of redundant space for embedding hidden information, making the research of audio information hiding techniques more valuable.

Traditional audio information hiding techniques can be roughly divided into two classes: time domain-based and transform domain-based techniques.

Time domain technique directly embeds the hidden information into the carrier signal in the time domain. It has large hiding capacity and easy to implement. However, directly modifying the carrier signal in the hiding process will inevitably cause distortion of the carrier signal, which will increase the difficulty of extracting the hidden information and make the imperceptibility be poor. The commonly used time domain-based techniques include the least significant bit (LSB) \cite{Dieu2013,Jadhav2016}, echo hiding \cite{Xiang2012,Hua2015} and spread spectrum \cite{Xiang2018,Xie2017} techniques.

Transform domain technique is to make the information hidden in the carrier's transform domain. It maps the carrier information to the transform domain and modifies some parameters in the transform domain to hide information, which can better resist the attack based on the various signal processing while maintaining the imperceptibility. However, mapping the audio signal to the transform domain requires a large number of signal processing operations, which results in high computation complexity. At the same time, the strong robustness is at the cost of reducing the hiding capacity. Therefore, the hiding capacity of transform domain technique is small. Commonly used transform domain-based techniques include phase coding \cite{Ngo2015,NGO2016}, discrete cosine transform (DCT) \cite{Zong2012,Jeyhoon2017} and discrete wavelet transform (DWT) \cite{Das2017,Avci2018} techniques.

In order to improve the hiding capacity and imperceptibility, this paper proposes to embed the hidden information to the audio signal by the private ASR model based on deep neural network (DNN) on the transmitting end and extract the hidden information by the private ASR model at the receiving end. We also perform several performance tests on the generated stego audios. Experiment results show that our proposed information hiding technique has good hiding capacity, imperceptibility and security. The contributions of this paper are as follows.

\begin{itemize}
\item \textbf{Novel hiding approach.} We propose a new audio information hiding technique based on the adversarial perturbations, which embeds and extracts the hidden information by the DNN-based ASR model.
\item \textbf{High hiding capacity.} The proposed technique embeds the hidden information in the form of a whole sentence with a hiding capacity of 48 character per second (cps).
\item \textbf{Well imperceptibility.} The value of perceptual evaluation of speech quality (PESQ) is 3.598 on average. People can barely perceive the perturbation.
\item \textbf{High security.} Four public models such as Google and IBM commercial ASR system are used to test the stego audio signals and experimental results show that these models are unable to extract the hidden information.
\item \textbf{A new backdoor.} The hidden information can be used as the specific trigger instruction to activate the model-intrinsic backdoor for DNN-based speech recognition models.
\end{itemize}

The remainder of this paper is organized as follows. Section~\ref{sec:relatedworks} introduces related works of traditional audio information hiding techniques. Section~\ref{sec:preliminary} indicates the preliminary knowledge. The proposed method and its working mechanisms are elaborated in Section~\ref{sec:proposedmethod}. The experimental results are reported in Section~\ref{sec:experimental}. Section~\ref{sec:application} demonstrates the application of our method and the intrinsic backdoor of DNN-based ASR models. Finally, we conclude in Section~\ref{sec:conclusion}.

\section{Related Works}
\label{sec:relatedworks}

The audio information hiding methods are mainly classed into time domain-based and transform domain-based methods. The time domain-based methods are characterized by low computation complexity and high hiding capacity, but poor robustness and imperceptibility. The transform domain-based methods usually have better robustness and imperceptibility, but the computation complexity is high and the hiding capacity is small. Several commonly used time domain-based and transform domain-based methods are introduced below.

\subsection{Time Domain-based Methods}

\subsubsection{Least Significant Bit (LSB) Method}

In the embedding phase, the LSB method replaces the least significant bits with the data bits of the hidden information. In the extraction phase, as long as the corresponding least significant bits are taken out, the embedded hidden information can be recovered. Dieu et al. \cite{Dieu2013} proposed an improved LSB method that is less sensible than the traditional LSB method, but it is at the cost of reducing the hiding capacity, and it is not robust. Jadhav et al. \cite{Jadhav2016} proposed an enhanced security audio information hiding technique that uses the top three most significant bits (MSBs) to determine the least significant bit (LSB) position of the hidden information. For example, when the top three bits are "100", the hidden information bit will be embedded in the 4th least significant bit. However, it still cannot solve the problem of poor robustness while improving security.

\subsubsection{Echo Hiding Method}

According to the auditory characteristics of human ear, if the weak signal appears in a short time (usually 0-200ms) after a strong signal in an audio, the weak signal will become inaudible. Echo hiding achieves the purpose of hiding information by introducing echoes into discrete audio signals, and various information can be represented by different echo delays. Xiang et al. \cite{Xiang2012} proposed a technique for embedding audio watermarks using echo hiding. The robustness and imperceptibility have been improved over previous work, but the hiding capacity has not been tested. Hua et al. \cite{Hua2015} proposed an audio watermarking scheme based on time-expanded echo. It uses the finite impulse response (FIR) filter based on convex optimization to obtain the optimal echo filter coefficients. This scheme improves the imperceptibility and robustness compared to the previous methods, but the hiding capacity has not been tested.

\subsection{Transform Domain-based Methods}

\subsubsection{DCT Domain Embedding Method}

DCT-based method obtains the DCT coefficients after DCT transform processing first, and then modifies the DCT coefficient for different positions to embed the hidden information into the audio. Finally, after the inverse discrete cosine transform operation, the stego audio signal is obtained. Zong et al. \cite{Zong2012} introduced an information hiding algorithm based on the energy difference between frequency bands. By calculating the difference between the energy average and the energy variation, the hidden information is embedded into the low frequency part of the DCT coefficients. The robustness is better. However, the calculation coefficients are too many and complicated and it is difficult to guarantee the correct rate of hidden information extraction. Jeyhoon et al. \cite{Jeyhoon2017} performed a DCT transform on each frame of the original audio signal and then selected the appropriate DCT coefficient band to embed the hidden information bits. The hiding capacity and robustness of this method are good, but the imperceptibility is slightly poor.

\subsubsection{DWT Domain Embedding Method}

Discrete wavelet transform (DWT) is a multi-scale multi-resolution technique that decomposes signals into different time-frequency components. Wavelet decomposition has a good match with the human ear's perceptual mental model. The difference between DWT-based and DCT-based method is that DWT modifies the DWT coefficient to embed the hidden information into the audio. Das et al. \cite{Das2017} proposed a method that hides both the hidden information and the key in the DWT coefficients. However, the paper did not test the hiding capacity and robustness of the stego audio. Avci et al. \cite{Avci2018} proposed a method by using the LSB method in the DWT domain. It has a good imperceptibility, but the hiding capacity is not high enough and no robustness experiments are performed.

A good information hiding algorithm should guarantee a large capacity with a good imperceptibility. However, the two indicators are usually contradictory. The large hiding capacity means that there is more hidden information can be embedded in the carrier audio. It will decrease the quality of the carrier audio, affect the imperceptibility, and increase the risk of being cracked. In this paper, we propose a new audio information hiding technique to balance imperceptibility and hiding capacity.

\section{Preliminaries}
\label{sec:preliminary}

\subsection{Automatic Speech Recognition}

Automatic Speech Recognition (ASR) \cite{Gruhn2011} is a cross-disciplinary applied research that transforms speech signals into corresponding texts through a process of recognition and understanding. Nowadays, speech recognition technology has been widely used in mobile devices, in-vehicle devices, robots and other scenes, and has played an increasingly important role in many fields such as search, manipulation, navigation, entertainment and so on.

Early speech recognition techniques are based on signal processing and pattern recognition methods. With the advancement of technology, machine learning methods are increasingly applied to speech recognition research, especially deep learning technology, which has brought profound changes to speech recognition research.

\begin{figure}[!htb]
\centering
\includegraphics[width=\columnwidth]{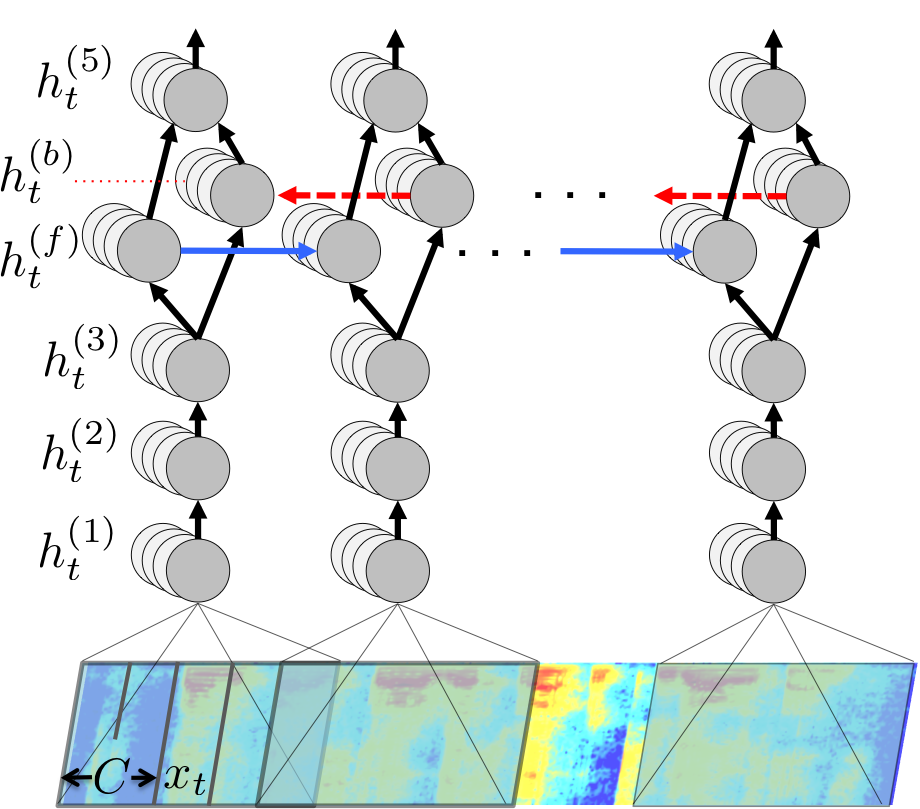}
\caption{The structure of DeepSpeech \cite{Hannun2014}.}
\label{fig:deepspeech}
\end{figure}

The structure of DeepSpeech \cite{Hannun2014} that we use is shown in Fig.~\ref{fig:deepspeech}, which is an open source ASR engine based on Baidu's deep speech research. The model is trained in deep learning techniques, which consists of five hidden layers \textit{ht$^{(1)}_t$-ht$^{(5)}_t$}. The bidirectional recurrent neural network (BiRNN) in the 4-th layer is the core of DeepSpeech and the loss function CTC-loss \cite{Hannun2017} is used to train the neural network.

\begin{figure*}[!htb]
\centering
\includegraphics{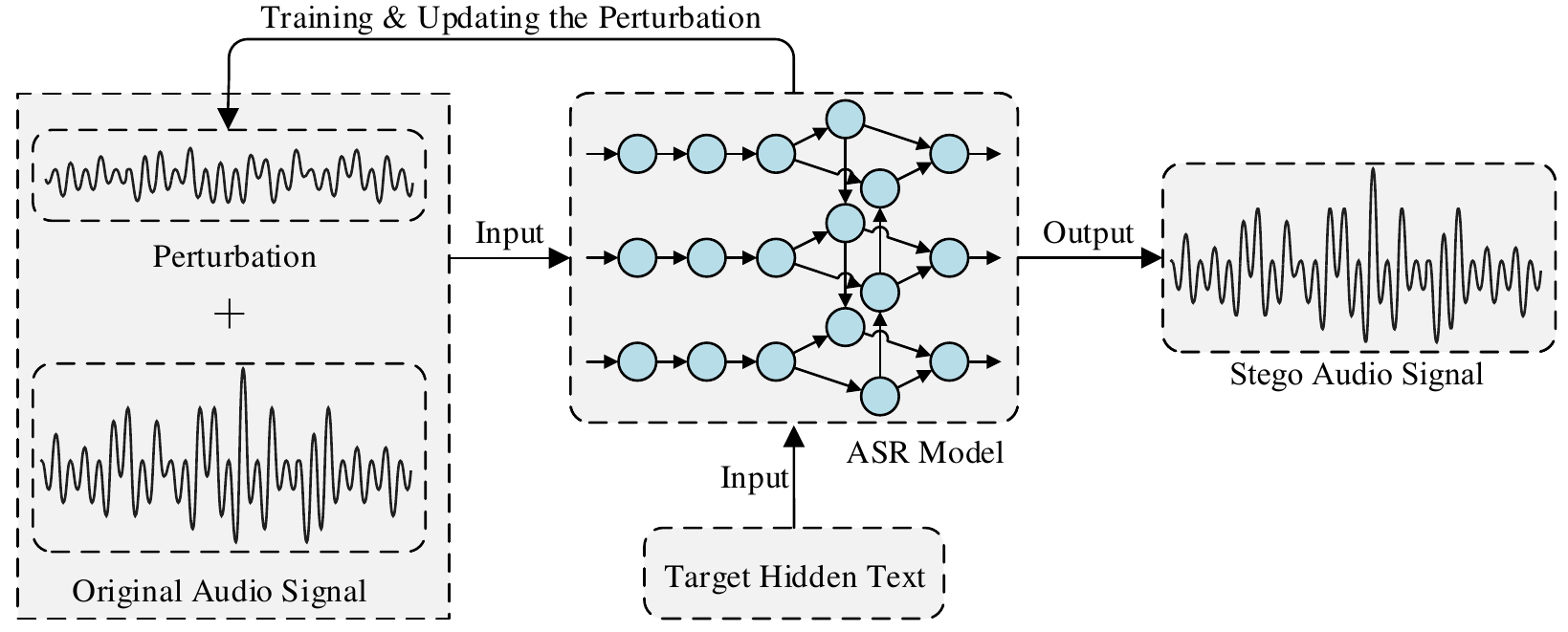}
\caption{The process of embedding the hidden information into audio signals.}
\label{fig:embed}
\end{figure*}

\subsection{Adversarial Examples}

Deep learning, especially neural networks, has shown great advantages in the fields of image recognition, speech processing, autonomous driving and medical diagnosis. In particular, the recognition ability of image recognition models has exceeded the accuracy of human eye. However, recent researches have shown that deep learning models are vulnerable to adversarial examples\cite{Szegedy2013}. Adversarial example is carefully designed by attackers to fool deep learning models. The difference between the adversarial examples and real examples is almost indistinguishable by the human eye, but it can cause the model to be misclassified.

The majority of adversarial example researches focused on generating adversarial examples against image recognition models \cite{Szegedy2013,Goodfellow2014,Kurakin2016,Papernot2016,Moosavi-Dezfooli2015,Carlini2017}. Szegedy et al. \cite{Szegedy2013} proposed an L-BFGS method that uses L2 distance norm square to constrain the perturbation to construct an image adversarial example. Although this method is stable and effective, the calculation is too complicated. Goodfellow et al. \cite{Goodfellow2014} added perturbations in the gradient direction causing the model to misclassify the resulting images. It is the simplest and fastest way to construct an adversarial example, called the fast gradient sign method (FGSM). However, as only one calculation is performed, the size of the perturbation cannot be well controlled. Carlini and Wagner \cite{Carlini2017} proposed an improved L-BFGS method, which is currently the most powerful attack method, called CW attack. It can perform targeted or non-targeted attacks effectively to misclassify the image recognition models.

However, recent researches have shown that speech recognition models are also vulnerable to adversarial examples \cite{Iter2017,Carlini2018,Taori2018a}. Iter et al. \cite{Iter2017} proposed  to use FGSM \cite{Goodfellow2014} to generate audio adversarial examples, which can misclassify the recognition result of the ASR system. However, the generated adversarial examples have loud noises through the human ear with a large distortion. Carlini et al. \cite{Carlini2018} proposed a method for generating audio adversarial examples against a white-box speech recognition model. It can produce very strong audio adversarial examples, resulting in a misclassification rate of up to 100\%. Taori et al. \cite{Taori2018a} combined genetic algorithm and gradient estimation targeted audio adversarial examples for black box ASR model. However, it can only generate a phrase consisting of two words with an attack success rate of only 35\%.

The vulnerability to adversarial examples threatens the security of DNN-based ASR models. However, we find an interesting characteristic that audio adversarial examples generated by the current generating methods have no transferability, that is, the audio adversarial examples generated by a specific model are not aggressive to other models. Therefore, we introduce this characteristic into the field of audio information hiding and use the deep learning model to generate audio adversarial examples to embed the hidden information.

\section{The Proposed Method}
\label{sec:proposedmethod}

The paper proposes a new technique based on the adversarial examples for audio information hiding, which embeds and extracts the hidden information by the private DNN-based ASR model. The technique is described in detail below. The proposed intrinsic backdoor of DNN-based model will be introduced in Section~\ref{sec:application}.

\subsection{Embedding Method}

The ASR model DeepSpeech acts as a private model owned only by the send end and the receive end, which functions like the grey box of embedding process in Fig.~\ref{fig:traditional} in the traditional information hiding method. The key idea is to take the original audio signal and hidden text as the input and obtain the stego audio signal after training the private model. For example, input an audio signal that is ``Good morning'' into the private model, and through training the model, the result of private model is finally recognized as ``hi, Siri''.

The process is shown in Fig.~\ref{fig:embed}. First, input the original audio signal \textit{X} and the hidden text \textit{t} into the ASR model. In training phase, the slight perturbation $\delta$ that needs to be added to the \textit{X} is constantly updated according to the result of the loss function. Finally, the generated stego audio signal $X+\delta$ can be recognized as the hidden text $t$ with a small perturbation $\delta$.

In order to recognize the audio as the hidden information, the CTC-loss is selected as the loss function of our method, which can output a probability for any text given an audio signal. The detail principle of CTC-loss can be found in \cite{Hannun2017}. Thus the stego audio can be trained using the CTC-loss to maximize the probability of the audio to be recognized as the hidden text.

In the meantime, under the premise that the audio is recognized as hidden text, the perturbation \textit{$\delta$} added to the original audio should be quieter than the original audio signal \textit{X}, hence its value should be smaller than the original one. To make a lower computational complexity, the L infinite norm $\|\delta\|_\infty$ (Eq.~\eqref{eq:norm}) is used to represent the magnitude of perturbation.

\begin{equation}
  \|\delta\|_\infty=\max \limits_{1\leq i \leq n}\ |\delta_i|.
\label{eq:norm}
\end{equation}

Converting the overall goal into an optimization problem is to minimize the $\|\delta\|_\infty$ in the case where the model \textit{C($\cdot$)} recognizes the speech \textit{(X+$\delta$)} as the target text \textit{t} (i.e., $C(X+\delta) =t$), that is,
\begin{equation}
\begin{aligned}
  &\min\;\|\delta\|_\infty\\
  &\ s.t.\ C(X+\delta) =t.
\end{aligned}
\end{equation}

Since the constraint $C(X+\delta) =t$ is non-linear, the gradient descent cannot be used to determine the convergence point of $\|\delta\|_\infty$. The constraint can be converted to minimizing the loss function $l(X+\delta, t)$, where the loss function $l(\cdot)$ is CTC loss and $l(X+\delta, t)$ indicates the magnitude of the CTC loss between the recognition result of $X+\delta$ and the target text $t$. Therefore, the optimization problem becomes two minimize problems. The formula is Eq.~\eqref{eq:min}:
\begin{equation}
  \min\;\|\delta\|_\infty\;\&\;\min\;l(X+\delta, t).
\label{eq:min}
\end{equation}

Hence how to combine the two constraints needs to be considered. For facilitating the application of the gradient optimizer, we separate them into two steps that keep iterating.

\begin{enumerate}
  \item Calculate the $\delta$ that meets the condition of $C(X+\delta)=t$ by applying gradient descent optimization to the loss function $l(X+\delta, t)$;
  \item Reduce the range of $\delta$ and clip it into the range.
\end{enumerate}

The two steps keep iterating until reaching the set threshold of iteration times. For the step 1, the gradient optimization of $\delta$ is performed by using Adam Optimizer to make the recognition result of $X+\delta$ close to the target text $t$ gradually. For the step 2, a threshold $\tau$ is set for $\delta$ to ensure the maximum fluctuation range of $\delta$ will not exceed the threshold. The two steps can be integrated to an iterative function Eq.~\eqref{eq:loss}:

\begin{equation}
  \left\{
    \begin{aligned}
    &\delta_0 = 0,\; X_0  =  X + \delta_0 \\
    &\delta_{N+1} = clip_{\delta, \tau}(\nabla_\delta l(X_{N},\; t)),\; X_{N+1}  =  X + \delta_{N+1} \\
    \end{aligned}
  \right.
\label{eq:loss}
\end{equation}

\begin{algorithm}
\renewcommand{\algorithmicrequire}{\textbf{Input:}}
\renewcommand{\algorithmicensure}{\textbf{Output:}}
\caption{Information Embedding Algorithm}
\label{alg1}
\begin{algorithmic}[1]
\REQUIRE Original audio signal $X$, Hidden text $t$
\ENSURE Stego audio signal $X'$
\STATE \textbf{Initialize:} $\delta-$an initial zero array with the same shape of $X$, $\tau-$the threshold of $\delta$, $N-$the max iteration times
\STATE $X' = X + \delta$
\FOR {$i = 0, 1, 2, \ldots, N$}
\STATE $//$Calculate the loss
\STATE $L=l(X', t)$
\STATE $//$Update $\delta$
\STATE $\delta\leftarrow$AdamOptimizer.$minimize(L,\delta)$
\STATE $\delta = clip(\delta, -\tau, \tau)$
\STATE $X' = X + \delta$
\IF {$C(X') == t $}
\label {code:10}
\STATE $//$Update the threshold $\tau$
\IF {$max(\delta)\leq\tau$}
\STATE $\tau = max(\delta)$
\ENDIF
\STATE $\tau = 0.8\cdot\tau$
\STATE $//$Save the last best result
\STATE $temp=X'$
\ENDIF
\label {code:18}
\ENDFOR
\RETURN $temp$
\end{algorithmic}
\end{algorithm}

The detailed algorithm is shown in Algorithm~\ref{alg1}. The function $clip(\delta, -\tau, \tau)$ sets the values of $\delta$ smaller than $\tau$ become $\tau$, and values larger than $-\tau$ become $-\tau$. As shown in Algorithm~\ref{alg1}, the determination of the threshold $\tau$ is shown in ~\ref{code:10} -~\ref{code:18}th rows. First, a large value of $\tau$ is given. Then after obtaining the minimized result $\delta$, the value of $\tau$ is reduced. The minimization process is repeated until reaching the number of iterations we set. Finally, the last best result will be returned.

\begin{figure*}[!htb]
\centering
\includegraphics{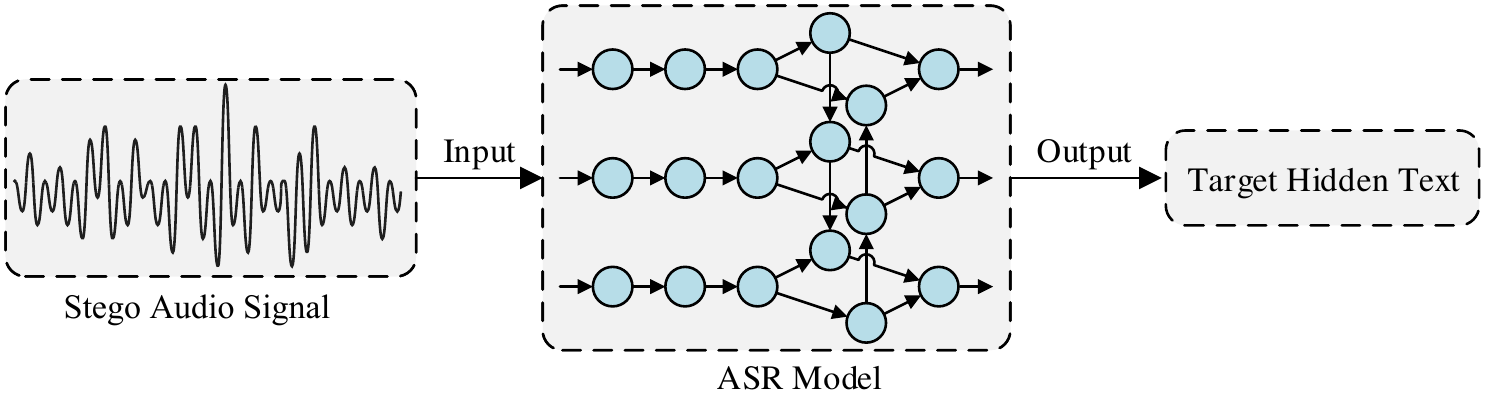}
\caption{The process of extracting the hidden information from stego audio signals.}
\label{fig:extract}
\end{figure*}

\subsection{Extracting Method}

Compared with traditional audio information hiding methods, our proposed method does not need to use any complicated algorithm to process the stego audio signal. As shown in Fig.~\ref{fig:extract}, the hidden text can be obtained by simply inputting the stego audio signal into the private ASR model to recognize. In order to ensure that other public ASR models cannot identify the hidden text, four state-of-art ASR models are used to extract the hidden text from the stego audios. The experimental results show that in addition to the private ASR model, all other public models cannot get any content related to the hidden text, the test results can be seen in Section~\ref{sec:experimental.c}.

\subsection{Backdoors}
Backdoors are typically activated under very specific conditions, which makes them unlikely to be activated and detected using random trigger inputs. This paper proposes to use the adversarial audio as the trigger input of DNN's intrinsic backdoor which is intrinsic for any DNN-based speech recognition models and not deliberately designed by the manufacturer. The intrinsic backdoor will be introduced in detail in Section~\ref{sec:application}.

\section{Experiments and Results}
\label{sec:experimental}

Although the requirements for information hiding are different in various scenarios, hiding capacity, imperceptibility, security and robustness are the main performance indicators of audio information hiding techniques \cite{Petitcolas1999,Simmons1998}. In addition, we perform a steganalysis to measure the probability of stego audio being discovered.

\begin{table}[!htb]
  \centering
  \caption{The specific hidden information in different groups. }
  \begin{tabular}{|c|c|m{0.5\columnwidth}<{\centering}|}
  \hline
  Group & Audio Range  & Hidden Information \\ \hline
  G1 & A00-A09 & be quiet \\ \hline
  G2 & A10-A19 & sing louder \\ \hline
  G3 & A20-A29 & close the door \\ \hline
  G4 & A30-A39 & the key is one one nine \\ \hline
  G5 & A40-A49 & call the police \\ \hline
  G6 & A50-A59 & happy birthday to you \\ \hline
  G7 & A60-A69 & be careful \\ \hline
  G8 & A70-A79 & bob is the spy \\ \hline
  G9 & A80-A89 & help me \\ \hline
  G10 & A90-A99 & see you at five pm \\ \hline
  \end{tabular}
  \label{tab:hidden_info}
\end{table}

\begin{table*}[!htb]
  \centering
  \caption{The hiding capacity of stego audios}
  \begin{tabular}{|c *{12}{|c}|}
  \hline
  \multicolumn{12}{|c|}{Proposed Method} & \multicolumn{1}{c|}{Method in \cite{Xiang2018}} \\ \hline
  Group	& G1 & G2 & G3 & G4 & G5 & G6 & G7 & G8 & G9 & G10 & Avg & \multirow{2}{*}{84 bps = 10.5 cps}\\ \cline{1-12}
  Capacity(cps) & 47.9 & 48.2 & 48.0 & 46.6 & 48.6 & 48.8 & 48.8 & 47.6 & 46.8 & 48.6 & 48.0 &\\ \hline
  \end{tabular}
  \label{tab:capacity}
\end{table*}

In order to test the performance of the proposed audio information hiding technique, 100 test audios (A00 - A99) are selected from the Mozilla common voice dataset \cite{Mozilla}, which are wav files with a length of 3 seconds, sampled at the rate of 16 kHz and quantized with 16 bits, to embed the hidden information. We divide these audios into 10 groups G1-G10. The specific information to be hidden in different groups is shown in Table~\ref{tab:hidden_info}. The stego audios are generated with tensorflow and DeepSpeech v0.1.0 version, and the evaluation indicators are obtained with MATLAB. The initial parameters we set in the experiments are as follows. The iteration times $N$ is 500, the initial $\delta$ is an array of 0 with the same shape of the audio signal, and the initial $\tau$ is set to 3000.

The performance of our method is compared with a spread spectrum-based audio information hiding method \cite{Xiang2018}. It obtains the DCT coefficients of audio first, then embeds the hidden information into the coefficients by using a group of orthonormal PN sequences. We have implemented this audio information hiding method in MATLAB using the original configuration in the \cite{Xiang2018}. The hiding capacity and imperceptibility are compared in the following evaluations.

\subsection{Hiding Capacity Analysis}

Hiding capacity, also known as the hiding rate, is the amount that hidden information can be embedded in the carrier signal per second. The traditional hiding methods hide information in the form of bits in audio, that is, the unit of hiding capacity is bit per second (bps). Our proposed method is directly hiding the information in the form of characters. The information extracted from the stego audio is a whole sentence, hence character per second (cps) is used as the unit of hiding capacity here.

\begin{figure}[!htb]
\centering
\includegraphics[width=\columnwidth]{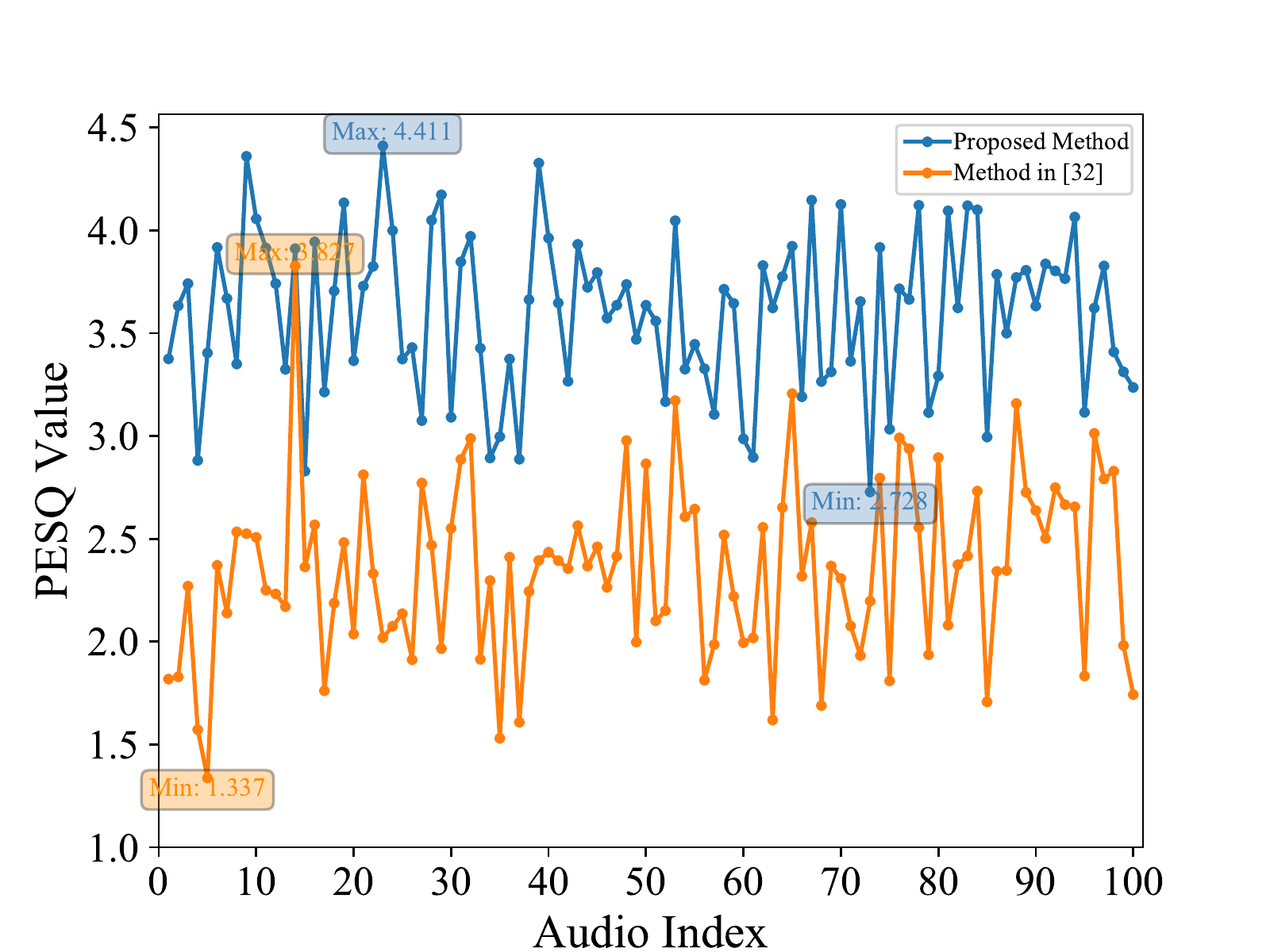}
\caption{The PESQ value of stego audios for the proposed method and the method in \cite{Xiang2018}.}
\label{fig:pesq}
\end{figure}

\begin{figure*}[!htb]
\centering
	\begin{minipage}{\columnwidth}
		\centering
		\includegraphics[width=\columnwidth]{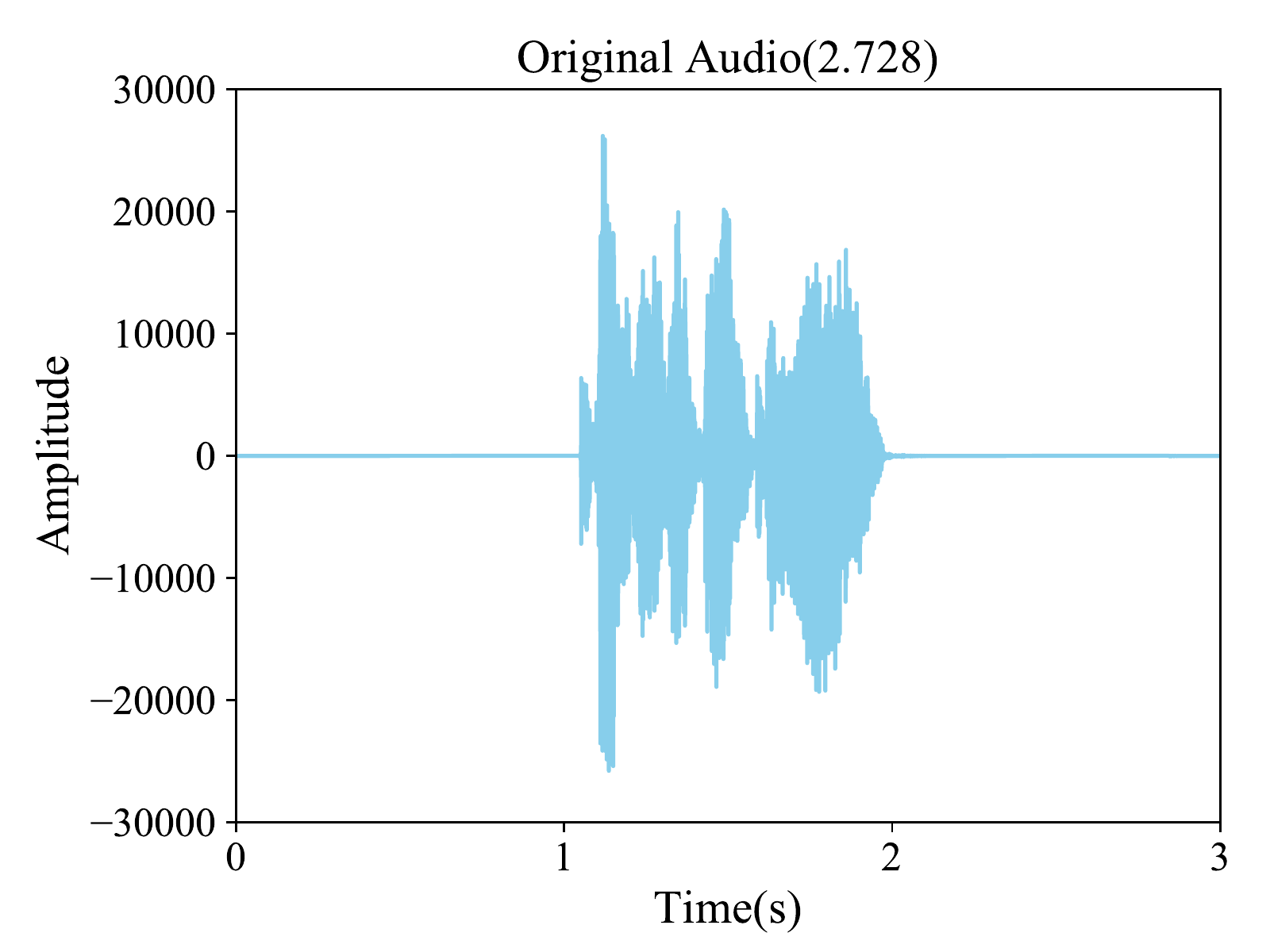}
        (a)
        \label{fig:a}	
	\end{minipage}
	\begin{minipage}{\columnwidth}
		\centering
		\includegraphics[width=\columnwidth]{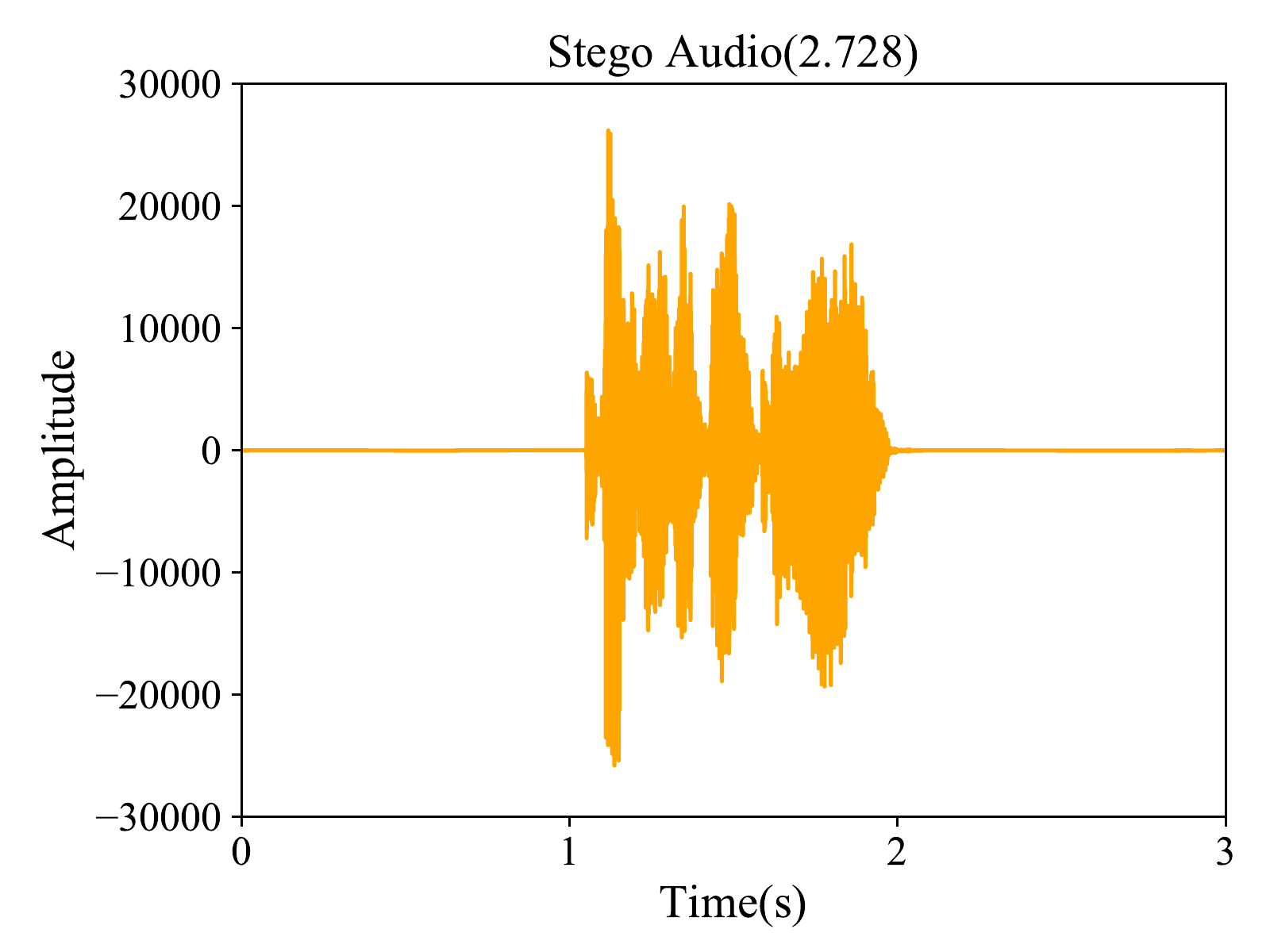}
        (b)
        \label{fig:b}
	\end{minipage}
	\begin{minipage}{\columnwidth}
		\centering
		\includegraphics[width=\columnwidth]{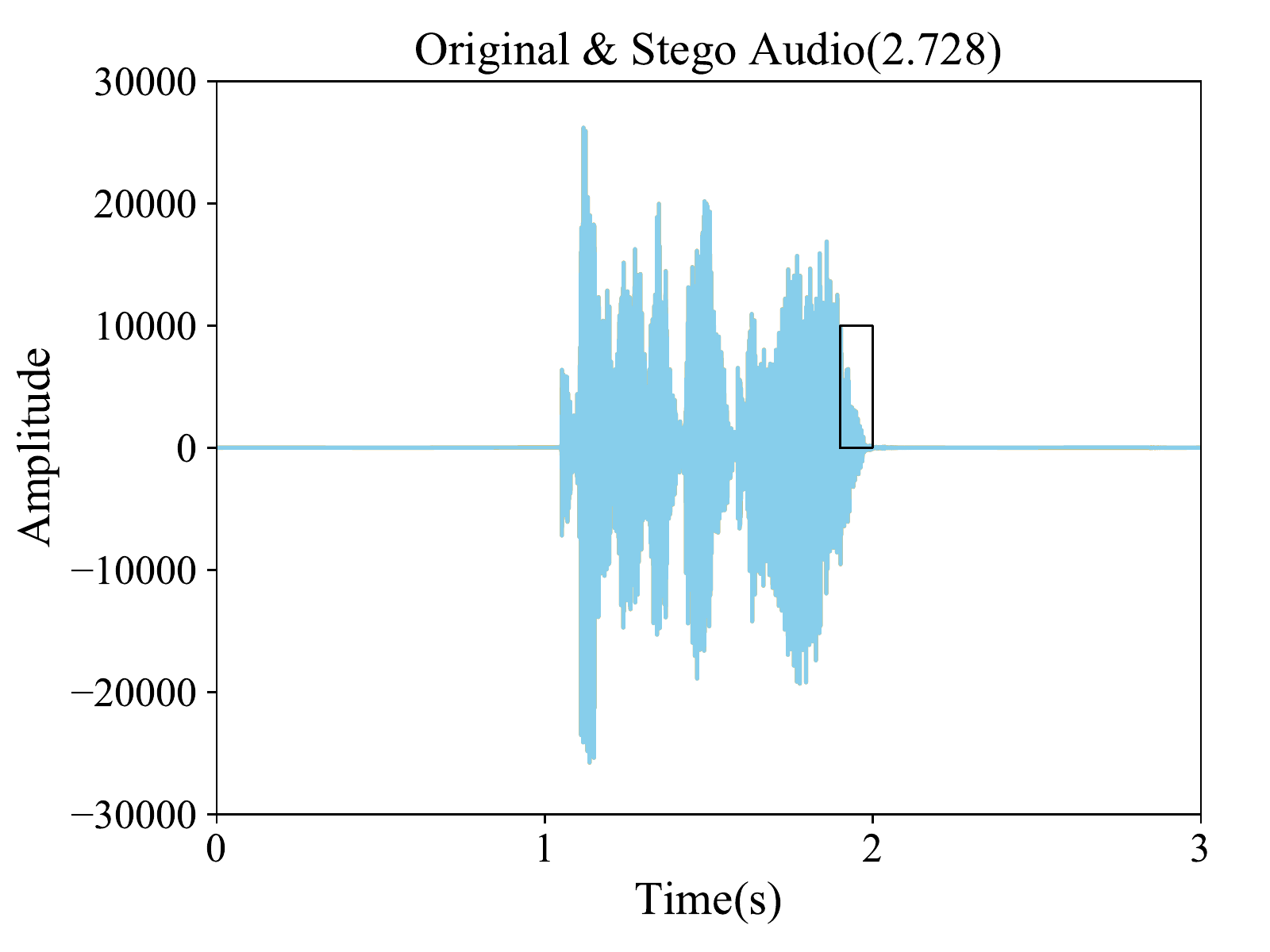}
        (c)
        \label{fig:c}
	\end{minipage}
    \begin{minipage}{\columnwidth}
		\centering
		\includegraphics[width=\columnwidth]{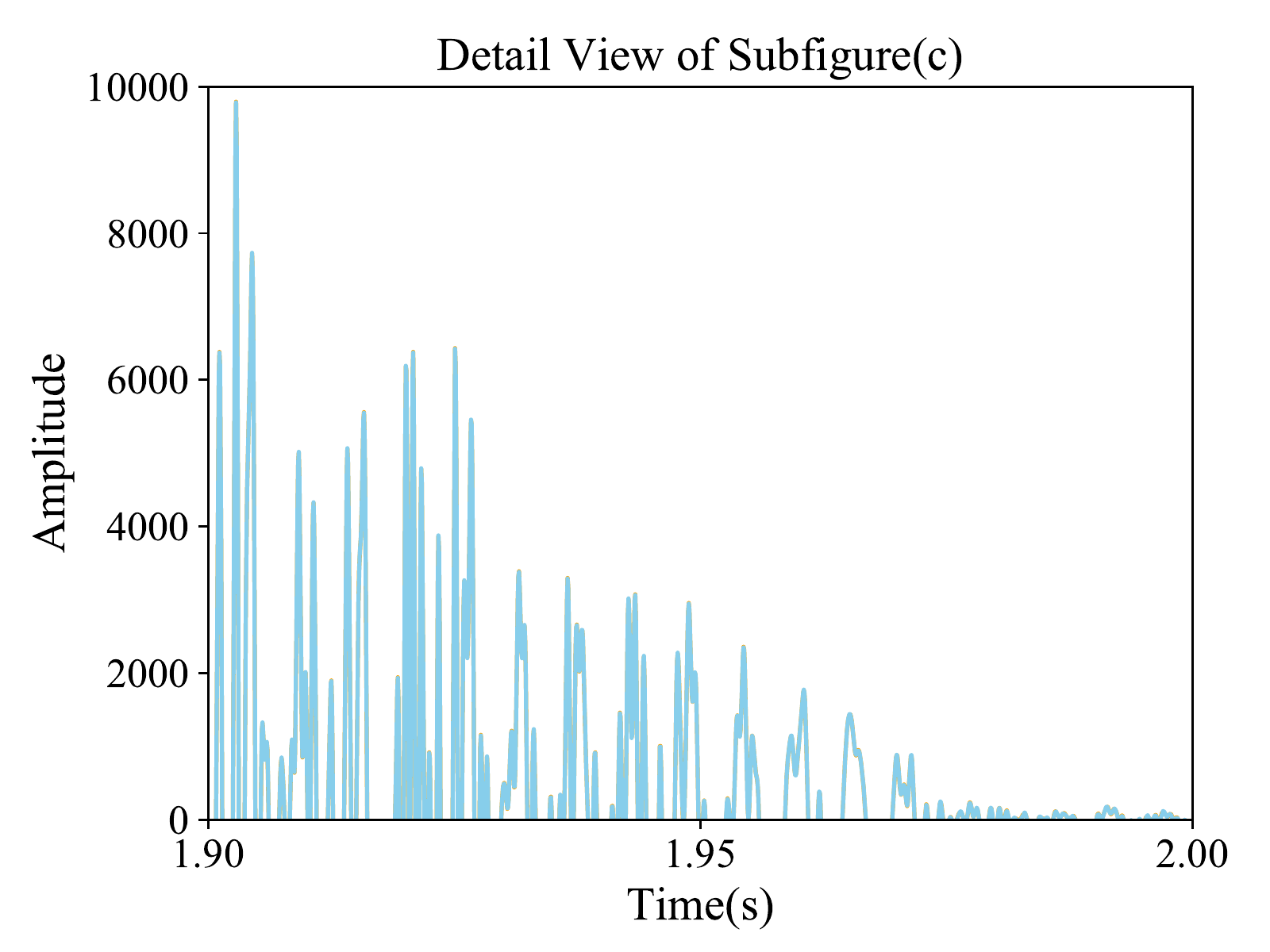}
        (d)
        \label{fig:d}
	\end{minipage}
\caption{The comparison of waveforms between original and stego audios.}
\label{fig:waveform}
\end{figure*}

Since the DeepSpeech model divides the audio signal into 50 frames per second when extracting speech features, which indicates that up to 50 characters can be recognized per second. Thus, the theoretical maximum capacity of this information hiding method is 50 cps. We conduct a hiding capacity test on the ten groups. The information for capacity analysis hidden for the audio signal per second is 10 consecutive "hide", separated by blanks. The experimental results are shown in Table~\ref{tab:capacity}, and the average hiding capacity is 48.0 cps. In the meantime, the hiding capacity of method in \cite{Xiang2018} is a fixed value of 84 bps. As 1 character equals to 8 bits, the capacity of \cite{Xiang2018} is 10.5 cps. Therefore, our proposed method has a higher hiding capacity.

\begin{table*}[htb]
  \centering
  \caption{The extracted text and extraction success rate of different ASR models }
  \begin{tabular}{|c|p{0.12\textwidth}<{\centering}|p{0.12\textwidth}<{\centering}|p{0.12\textwidth}<{\centering}|p{0.12\textwidth}<{\centering}|p{0.12\textwidth}<{\centering}|}
  \hline
  \multirow{2}{*}[-7pt]{Group} &\multicolumn{2}{c|}{Model Internal Security} &\multicolumn{3}{c|}{Model External Security} \\ \cline{2-6}
   & \tabincell{c}{DeepSpeech\\ v0.1.0} & \tabincell{c}{DeepSpeech\\ v0.2.0} & Google Cloud & IBM Watson & iFlytek \\ \hline
  G1 & 100\% & 0\% & 0\% & 0\% & 0\% \\ \hline
  G2 & 100\% & 0\% & 0\% & 0\% & 0\% \\ \hline
  G3 & 100\% & 0\% & 0\% & 0\% & 0\% \\ \hline
  G4 & 100\% & 0\% & 0\% & 0\% & 0\% \\ \hline
  G5 & 100\% & 0\% & 0\% & 0\% & 0\% \\ \hline
  G6 & 100\% & 0\% & 0\% & 0\% & 0\% \\ \hline
  G7 & 100\% & 0\% & 0\% & 0\% & 0\% \\ \hline
  G8 & 100\% & 0\% & 0\% & 0\% & 0\% \\ \hline
  G9 & 100\% & 0\% & 0\% & 0\% & 0\% \\ \hline
  G10 & 100\% & 0\% & 0\% & 0\% & 0\% \\ \hline
  Average Success Rate & 100\% & 0\% & 0\% & 0\% & 0\% \\ \hline
  \end{tabular}
  \label{tab:security}
\end{table*}

\subsection{Imperceptibility Analysis}

The evaluation methods of audio imperceptibility can be classed into subjective evaluation and objective evaluation. Subjective method evaluates the quality of audio based on human hearing. It is consistent with people's perception of audio quality. However, the shortcomings are time-consuming and labor-intensive, lack of flexibility, poor repeatability and stability. Objective evaluation utilizes machine to discriminate audio quality automatically. It gives audio quality evaluation results in a convenient and fast way without subjective influence. In this paper, perceptual evaluation of speech quality (PESQ) is used to perform imperceptible analysis of audio signals. PESQ is an objective mean opinion score (MOS) value evaluation method provided by ITU-T Recommendation P.862, which uses the stego audio to compare with the original audio. In general, the score is between 1.0 and 4.5. The worse the speech quality, the lower the score.

In order to evaluate the imperceptibility of the proposed method, we select a recently proposed audio information hiding method \cite{Xiang2018} for comparison. 100 stego audios are generated using the two embedding methods with the hidden text in Table~\ref{tab:hidden_info}. The tested PESQ value results are shown in Fig.~\ref{fig:pesq}, the average PESQ value of our proposed method is 3.598, while the method in \cite{Xiang2018} is 2.351.

The audio with the lowest PESQ value is selected to analyze the waveform of the audio signals. The waveforms are shown in Fig.~\ref{fig:waveform}. In general, the smaller the difference between original audio and stego audio, the larger the PESQ value, that is, the imperceptibility is better. Fig.~\ref{fig:waveform} (a) is the waveform of original audio and Fig.~\ref{fig:waveform} (b) is the waveform of stego audio. In order to find the difference between the two waveforms more intuitively, we combine the two into Fig.~\ref{fig:waveform} (c). It can be found that the two audios are basically overlapped completely and have no difference. Thus, a small part, which is the black box in Fig.~\ref{fig:waveform} (c), is enlarged for more detailed observation in Fig.~\ref{fig:waveform} (d). By observing the Fig.~\ref{fig:waveform} (d), even if enlarged 30 times, the difference between them is still small, which means that our method has good imperceptibility.

\subsection{Security Analysis}
\label{sec:experimental.c}

The security of audio information hiding refers to the ability that the hidden information cannot be extracted by the attacker. In this paper, the original model (DeepSpeech v0.1.0) is used as the private model.

\subsubsection{Model Internal Security Analysis}
As the private model acts like the key in traditional hiding methods, the key space security should be evaluated, which is defined as model internal security in our proposed method. The model internal security analysis is to find out if the model will output the same result while the weights of model are different. We evaluate the security by comparing the extraction success rate from the private model to its upgraded version model DeepSpeech v0.2.0. The two DNN models have different neuron weights while holding the same neural network structure. The extracting results are shown in Table~\ref{tab:security}.

\subsubsection{Model External Security Analysis}
The model external security analysis is to find out if the model will output the same result while the whole model structure and parameters are different. We evaluate the model external security by comparing the extraction success rate from the private model to other ASR models. Three public commercialized ASR platform services Google Cloud \cite{Google}, IBM Watson \cite{IBM} and iFlytek \cite{iFlytek} Speech-to-Text are selected to extract the hidden information in different groups. The extraction success rates are shown in Table~\ref{tab:security}.\\

From the above results, it can be seen that only the private model can extract the hidden information. Even the same model cannot extract hidden information after the model parameters are updated (i.e., DeepSpeech v0.2.0). In addition, according to the specific extraction information during the experiment, only the information related to the original audio can be obtained for public models. Any content related to the hidden text cannot be obtained at all. Therefore, the security of this audio information hiding method is high.

\subsection{Robustness Analysis}

The robustness of audio information hiding refers to the ability of the stego audio to remain hidden text that can be completely extracted after suffering some modification or transformation. In order to test the robustness of the algorithm, the above 10 stego audio groups are processed as follows:

\begin{table*}[!htb]
  \centering
  \caption{The extraction success rate after 4 signal processing methods}
  \begin{tabular}{|c|c|c|c|c|}
  \hline
  Group & \multicolumn{1}{c|}{White Gaussian Noise} & \multicolumn{1}{c|}{Resampling} & \multicolumn{1}{c|}{Lowpass Filtering} & \multicolumn{1}{c|}{Echo Interference} \\ \hline
  G1 & 0\% & 50\% & 0\% & 0\% \\ \hline
  G2 & 0\% & 30\% & 0\% & 0\% \\ \hline
  G3 & 0\% & 70\% & 0\% & 0\% \\ \hline
  G4 & 0\% & 70\% & 0\% & 0\% \\ \hline
  G5 & 0\% & 50\% & 0\% & 0\% \\ \hline
  G6 & 0\% & 60\% & 0\% & 0\% \\ \hline
  G7 & 0\% & 20\% & 10\% & 10\% \\ \hline
  G8 & 0\% & 60\% & 0\% & 0\% \\ \hline
  G9 & 0\% & 40\% & 0\% & 0\% \\ \hline
  G10 & 0\% & 10\% & 0\% & 0\% \\ \hline
  Average Success Rate & 0\% & 46\% & 1\% & 1\% \\ \hline
  \end{tabular}
  \label{tab:robustness}
\end{table*}

\begin{enumerate}
  \item Add Gaussian white noise. A Gaussian white noise with a signal-noise ratio (SNR) of 20 dB is added to the stego audio signals;
  \item Resampling attack. Up-sampling the stego audio signals. Up-sampling: First, the stego audio signal is resampled by 2 times the original sampling rate, and then restored to the original sampling rate;
  \item Low-pass filtering: The Butterworth low-pass filter with a 2-order cutoff frequency of 6 kHz is processed for stego audio signals;
  \item Echo interference: Add an echo with a 50\% attenuation rate and a delay of 30ms in the stego audio signals.
\end{enumerate}

As shown in Table~\ref{tab:robustness}, except the resampling attack, the stego audio signals have lost the hidden text after being processed by these methods. The experimental results show that the robustness of our proposed hiding technique is not good. Therefore, in order to enable the receiving end to extract the hidden text successfully, the stego audio signals can only be transmitted in a lossless propagation, for example, to upload the audio file.

\subsection{Steganalysis}
Steganalysis is a technique against information hiding for detecting whether there is hidden information in data. According to the relationship between feature extraction and embedding algorithm, it can be classed into two steganalysis technologies, which are called as special steganalysis and general steganalysis technique, respectively. The special steganalysis generally targets a certain type of information hiding method. According to the statistical analysis of the data, it uses the difference between the statistical features to design the corresponding steganalysis algorithm. The general steganalysis generally aims at multiple types of hiding methods, which is more universal and has more practical value. It extracts some features of the data to form a feature vector set, which is then trained by neural network, clustering algorithm or other methods to construct a detection model to analyze the hidden information.

In this paper, we use a recently proposed general steganalysis method that takes the quantified modified DCT (QMDCT) coefficients matrix as the input of a convolutional neural network (CNN) \cite{WangYYZX18} to analyze if the audio signal have been embedded hidden information. The CNN model is trained with the default configuration and dataset in \cite{WangYYZX18}. Then the 200 original and stego audios are analyzed by the model. The results show that 100\% stego audios can be recognized as being embedded hidden information. However, 80\% original audios are misidentified as well, which means the results can not be trusted. Therefore, the current mainstream audio steganalysis algorithm is not accurate enough for our proposed DNN-based audio information hiding method.

\section{DNN-intrinsic Backdoor}
\label{sec:application}
This paper proposes a new audio information hidden technique, which also can be used to activate an intrinsic backdoor of DNN-based ASR models.

The intelligent speakers, which are the core part of the intelligent home control system market, have been developed rapidly. In the meanwhile, as the most basic component of intelligent speakers, automatic speech recognition (ASR) service is widely integrated into them like Amazon Echo~\cite{Echo}, Google Home~\cite{Home}, Xiaomi AI speaker~\cite{Xiaomi}, and Tmall Genie~\cite{Tmall}, etc.

\begin{figure*}[!htb]
\centering
\includegraphics[width=2\columnwidth]{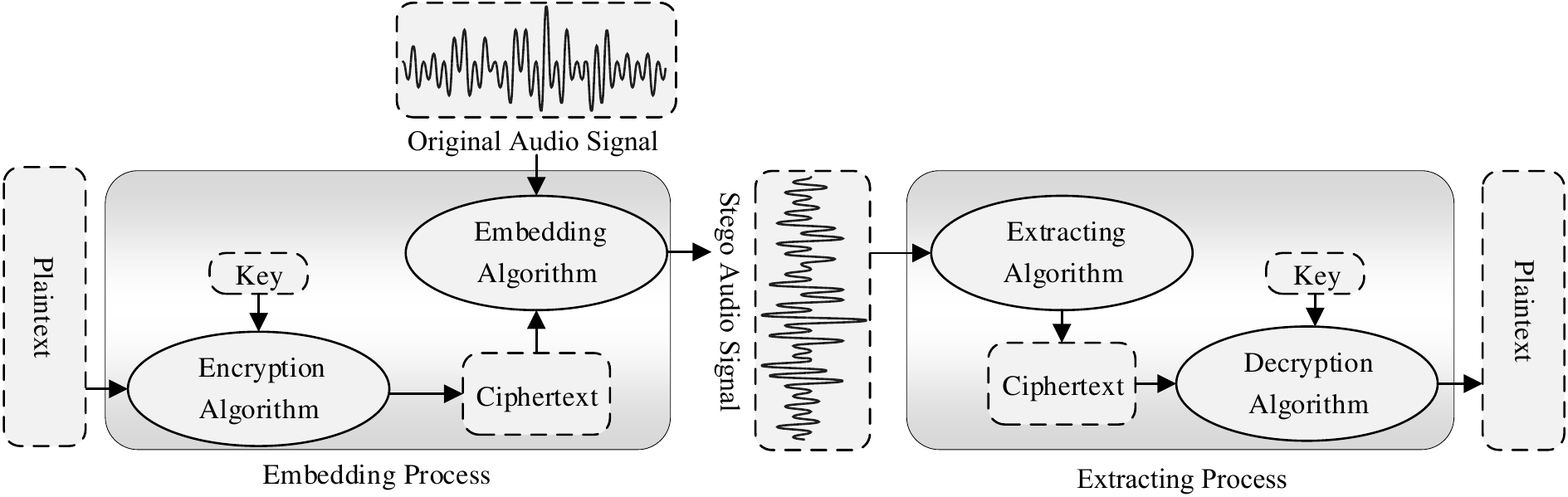}
\caption{The embedding and extracting process of traditional information hiding methods.}
\label{fig:traditional}
\end{figure*}

When deploying an information hiding technique on IoT devices like intelligent speakers, the overheads should be considered because of the limited resources such as CPU, memory, and battery power.

Fig.~\ref{fig:traditional} shows the whole process of traditional information hiding methods. In the extracting process, the encrypted information of the stego audio needs to be decrypted after being extracted. However, classic cryptographic security solutions incur expensive overheads, which is unacceptable on resource-constrained IoT devices. On the contrary, as shown in Fig.~\ref{fig:extract}, the information hiding method we propose does not need the decryption unit and key storage. The hidden information can be obtained by simply inputting the stego audio into the ASR model to recognize, which means the overhead is negligible. Thus it can be an appropriate solution for audio information hiding in resource-constrained IoT devices like intelligent speakers.

\begin{figure}[!htb]
\centering
\includegraphics[width=\columnwidth]{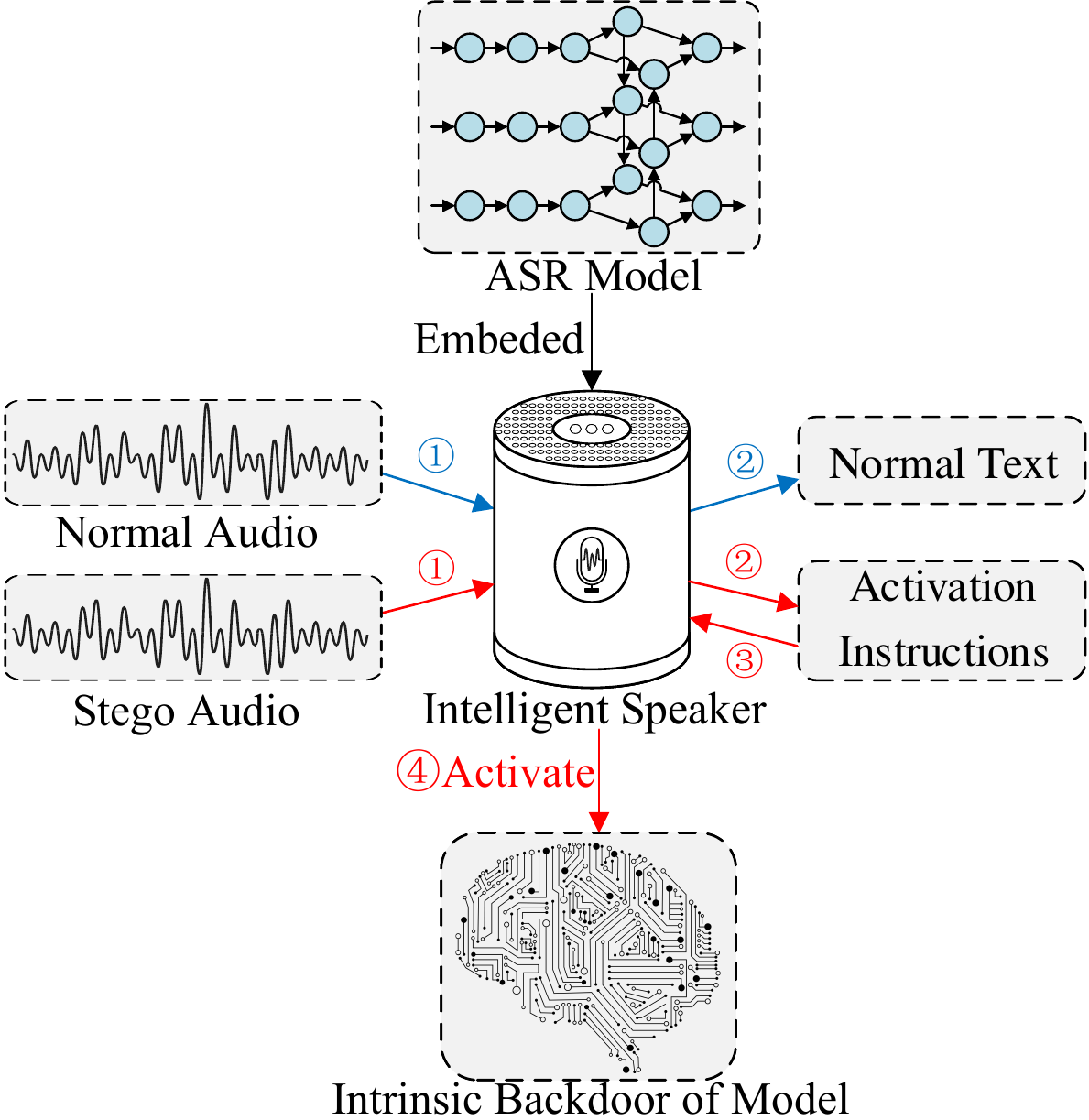}
\caption{The recognition process of the DNN-based intelligent speaker. The blue lines indicate the process of recognizing a normal audio. The red lines indicate the process of activating the intrinsic backdoor of ASR model by recognizing the stego audio.}
\label{fig:backdoor}
\end{figure}

However, the hidden information can be illegally used by attackers to activate a backdoor of the DNN-based ASR model. Fig.~\ref{fig:backdoor} shows an example of the recognition process of the intelligent speaker. As shown in Fig.~\ref{fig:backdoor}, the intelligent speaker, which embeds with a DNN-based ASR model, can recognize the normal audio as a normal sentence like the process indicated by the blue lines in the figure. However, when a stego audio with the hidden information of activation instruction is sent to the intelligent speaker, it will be recognized as a command to activate the backdoor of DNN model like the process indicated by the red lines. Later on, once the backdoor of the intelligent speaker is activated, attackers can obtain the control of all intelligent IoT devices in your home. For example, open the door, query the position of children's watch, and control curtain, air conditioner, TV, etc. At the same time, your personal privacy information may be leaked through these devices, which can be used to carry out illegal activities or cause damage to your property.

\section{Conclusions}
\label{sec:conclusion}
The paper proposes a novel technique for audio information hiding based on adversarial examples, which takes the original audio signal as input and obtains the stego audio through the training process of the private ASR model. According to experimental results, the generated stego audio signal has a hiding capacity of 48.0 cps with good imperceptibility, which is difficult for the human ear to perceive the difference between the original audio signal and the stego audio signal. Besides, The hidden text in stego audio signal can only be extracted by the private ASR model. Without knowing the internal parameters and structure of the private model, the public model can only extract the original text. Therefore, the security of our proposed audio information hiding method is high. In addition, our proposed adversarial audio brings serious threats to DNN-based ASR models.

However, our proposed new audio information hiding technique is not robust enough. At current stage, the stego audio signals can only be transmitted in a lossless propagation. We expect to provide a new solution for audio information hiding, and gradually address the shortcoming in further research.

\bibliographystyle{plain}
\bibliography{bibliography}

\end{document}